%% file: ups.tex
\documentclass{elsart}
\usepackage{epsf}
\usepackage{pst-node}
\usepackage{graphics}
\usepackage{amssymb}
\begin{document}             

\begin{frontmatter}

\title{\mbox{Measurement of the $\boldmath\Upsilon$ Production Cross Section}\\
 \mbox{in 920\,\,GeV Fixed-Target Proton-Nucleus Collisions}
}

\begin{center}
                       {\bf The HERA-B collaboration}
\end{center}
\input{authors_epjc.tex}
\vspace*{-3ex}
\begin{abstract}
% Text of abstract
The proton-nucleon cross section ratio 
$R=\mbox{Br}(\Upsilon\to{}l^+l^-)\cdot{d\sigma(\Upsilon)/dy}|_{y=0}\,/\,{\sigma(J/\psi)}$
has been measured 
with the HERA-B spectrometer 
in fixed-target
proton-nucleus collisions at 920~GeV proton beam energy 
corresponding to a proton-nucleon cms energy of $\sqrt{s}=41.6$~GeV.
  The combined results for the %$\Upsilon$~
decay channels
$\Upsilon\to e^+e^-$ and $\Upsilon\to\mu^+\mu^-$
  yield a ratio $R=(9.0 \pm 2.1)\cdot 10^{-6}$.
The corresponding
$\Upsilon$~production cross section per nucleon at mid-rapidity ($y=0$) has been determined
to be
$\mbox{Br}(\Upsilon\to{}l^+l^-)\cdot{d\sigma(\Upsilon)/dy}|_{y=0}= 4.5 \pm 1.1 $~pb/nucleon.
\end{abstract}

\date{ }
\begin{keyword}

$\Upsilon$~mesons
\sep
cross section
\sep
proton-nucleus collisions
\sep
$\sqrt{s}=41.6$~GeV

\PACS
{13.20.Gd} % Decays of $J/\psi$, $\Upsilon$, and other quarkonia
\sep
{13.85.Qk} % Inclusive production with identified leptons, photons, or other nonhadronic particles
\sep
{14.40.Gx} % Mesons with $S=C=B=0$, mass~$>2.5$~GeV (including quarkonia)
\sep
{24.85.+p} % Quarks, gluons, and QCD in nuclei and nuclear processes
\end{keyword}

\end{frontmatter}

%%upright Greek letters (example below: upright ``mu'')
\newcommand{\greeksym}[1]{{\usefont{U}{psy}{m}{n}#1}}
\newcommand{\umu}{\mbox{\greeksym{m}}}
\newcommand{\udelta}{\mbox{\greeksym{d}}}
\newcommand{\uDelta}{\mbox{\greeksym{D}}}
\newcommand{\uPi}{\mbox{\greeksym{P}}}
%%%%%%%%%%%%%%%%%%%%%%%%%%%%%%%%%%%%%%%%%%%%%%%%%%%%%%%%%%%%%

%=======================
 \section{Introduction}
%=======================

In recent years, there has been a rapid development of models 
describing quarkonium, especially charmonium,
production, with great success both in the high- and low-energy
range (see e.g.\ review \cite{vogt,QWG} and references therein). 
These developments are driven by the
available measurements of charmonium production, and the implication 
of a possible suppression of charmonium production 
as an indicator for a Quark-Gluon Plasma (QGP) \cite{QGP1,QGP2}.
Bottomonium production represents a natural field for testing
the predictions of these models~\cite{vogt}. 
Measurements of the $\Upsilon$~production cross 
section\footnote{Throughout this paper we refer to the sum of 
$\Upsilon(1S)+\Upsilon(2S)+\Upsilon(3S)$ cross sections
as the $\Upsilon$~cross section.} have been performed
by many experiments \cite{badier}--\cite{cdf}
in the wide range of the proton-nucleon centre-of-mass (cms) energy $\sqrt{s}$ of 19 to 1800~GeV,
and with targets ranging from proton ($A=1$) to platinum ($A=195$)
with both proton and antiproton beams.

Usually, the result of bottomonium production measurements is  
presented as the product of the differential cross section at 
mid-rapidity $d\sigma(\Upsilon)/dy\big|_{y=0}$
times branching ratio~\cite{vogt}.
  However, the acceptance of most of the fixed-target experiments
which have so far measured bottomonium production, is not in the
region of central collisions, $y=0$, so that systematic
effects in determining the total cross section can be substantial.
  More specifically, precisely in the energy region of HERA-B,
the available results \cite{moreno,McGaughey}
disagree by about a factor of two
(see Table~\ref{tab-exper}%
%%%%%%%%%%%%%%%%%%%%%%%%%%%%%%%%%%%%%%%%%%%%%%%%%%%%%%%%%%%%%%%%%%%%%%%%%%%%%%%%%%%%%%%%%%%%%
\footnote{Both papers \cite{moreno,McGaughey} quote the differential cross section
in terms of the Feynman scaling variable $x_F$ which we transform to rapidity $y$. 
 The two quantities are related via
$d\sigma(\Upsilon)/dx_F\big|_{x_F=0}=F(\sqrt{s})\cdot d\sigma(\Upsilon)/dy|_{y=0}$,
where $F(\sqrt{s})$ is a coefficient which depends on the
(measured) transverse momentum distribution of the produced $\Upsilon$ mesons.
Its numerical value is $1.98\pm0.03$ and $2.12\pm0.03$ for $\sqrt{s}=38.8$~GeV and
$\sqrt{s}=41.6$~GeV, respectively.}%
%%%%%%%%%%%%%%%%%%%%%%%%%%%%%%%%%%%%%%%%%%%%%%%%%%
) which cannot be explained by a large nuclear suppression~\cite{e866-alpha}.
  HERA-B covers the region of mid-rapidity, implying less uncertainty
in the determination of the total cross section,
and can thus contribute to the
know\-ledge of $\Upsilon$~production, despite its rather small sample size.
  Furthermore, both $\Upsilon\to\mu^+\mu^-$ and $\Upsilon\to e^+e^-$
decay channels are measured simultaneously, providing an additional statistically
independent cross check.

%==============================================================================
% Table 1
%==============================================================================
\begin{table}
\begin{center}
\caption{Summary of the available measurements of the $\Upsilon$~production cross section 
  in $p$A collisions near $\sqrt{s}$ = 41.6~GeV.
    The published result of \cite{McGaughey} refers to the
  production of $\Upsilon$(1S) only and has been rescaled  using Eq.\,(\ref{e605-fractions}) to also include
  $\Upsilon$(2S) and $\Upsilon$(3S).
In the case of E771 \cite{e771}, the published value has been corrected
to include the $\Upsilon$(3S) state and the coefficient $\Delta y_{\rm eff}$ (see Sect.\,\ref{sect-syst}).
Overall normalization uncertainties of 15\% (Ref.\,\cite{moreno}) and 10\% (Ref.\,\cite{McGaughey})
 have been included.
}
%%%%%%%%%%%%%%%%%%%%%%%%%%%%%%%%%%%%%%%%%%%%%%%%%%%%%%%%%%%%%%%%%%%
\label{tab-exper}
\begin{tabular}{|c|c|c|c|}
\hline
$\sqrt{s}$, & $\mbox{Br}(\Upsilon\to{}l^+l^-) \cdot d\sigma(\Upsilon)/dy\big|_{y=0}$, & Tar- & Expe- \\[-1ex] 
 GeV     & pb/nucleon                                   & get  & riment \\
\hline
38.8  & $2.11\pm 0.33$     & Cu & E605 \cite{moreno} \\
38.8  & $2.31\pm 0.38$      & Be & E605 \cite{yoshida} \\
38.8  & $4.7 \pm 0.5$      & D  & E772 \cite{McGaughey}\\
38.8  & $7.7 \pm 3.2$      & Si & E771 \cite{e771}\\
\hline
\end{tabular}
\end{center}
\end{table}
%==============================================================================

%=======================
\section{Measurement method}
%=======================

  We determine the $\Upsilon$~production cross section
by comparing the relative 
yields of $\Upsilon$ and $J/\psi$ production, and normalizing to
the known $J/\psi$ cross section: % (\cite{e789,e771-jpsi}).
\begin{eqnarray}
\hspace*{-8mm}
\mbox{Br}(\Upsilon\to l^+l^-)\cdot{d\sigma\over dy}(\Upsilon)\Big|_{y=0}=
\mbox{Br}(J/\psi\to l^+l^-) 
% \nonumber \\ 
\cdot \,\sigma(J/\psi)
   \cdot {N(\Upsilon)\over N(J/\psi)}
   \, {\varepsilon(J/\psi) \over {\varepsilon(\Upsilon)}}
   \, {1\over\Delta y_{\rm eff}}
\hspace*{5mm}
\label{basic}
\end{eqnarray}
and in addition, we define the ratio of the mid-rapidity $\Upsilon$ cross section 
to the total $J/\psi$ cross section as
\begin{equation}
R_{J/\psi} \equiv {{\mbox{Br}(\Upsilon\to l^+l^-)\cdot
{d\sigma(\Upsilon)/dy}}|_{y=0} \over \sigma(J/\psi)}\,\,.
\label{eq-rjpsi}
\end{equation}

Here, $\sigma(J/\psi)$ is the $J/\psi$ production cross section, and
 $N(\Upsilon)$ and $N(J/\psi)$ are the numbers of observed $\Upsilon$
and $J/\psi$ decays, respectively. 
The branching ratio $J/\psi\to l^+l^-$ is taken as the average of
the most recent values for $J/\psi\to e^+e^-$ and $J/\psi\to \mu^+\mu^-$~\cite{pdg},
  $\varepsilon(J/\psi)/\varepsilon(\Upsilon)$ is the ratio of 
$J/\psi$ and $\Upsilon$~trigger and reconstruction efficiencies
determined from Monte Carlo simulations,
and $\Delta y_{\rm eff}$ is the coefficient relating the full and 
differential $\Upsilon$ cross section at mid-rapidity $y=0$:
$\sigma(\Upsilon) = \Delta y_{\rm eff} \,\, d\sigma(\Upsilon)/dy\big|_{y=0}$
(see Sect.\,\ref{sect-syst}).
  In this way the result is independent of the luminosity determination, and 
systematic uncertainties due to 
absolute trigger efficiencies cancel out to a large extent, since only relative 
efficiencies between $\Upsilon$ and $J/\psi$ and relative acceptances enter the final result.
  For $\sigma(J/\psi)$ at $\sqrt{s}=41.6$~GeV we use the value 
\begin{equation}
 \sigma_{pN}(J/\psi)=502 \pm 44\,{\mbox{nb}\over\mbox{nucleon}}
\label{jpsi-reference}
\end{equation}
obtained 
from a global fit \cite{jpsi-reference} of $J/\psi$ production data%
%%%%%%%%%%%%%%%%%%%%%%%%%%%%%%%%%%%%%%%%%%%%%%%%%%%%%%%%%%%%%%%%%%%%%%%%%%%%%%%%%
\footnote{The global fit includes our own measurement of the $J/\psi$ cross section 
$
% \sigma_{pN}^{HERA-B}(J/\psi)=
663 \pm 74 \pm 46\,{\mbox{nb}/\mbox{nucleon}}$
\cite{herab-jpsi}. 
We prefer to normalize the present measurement to the value given by
the global fit since the fit provides a complete summary of all
available measurements.}%
%%%%%%%%%%%%%%%%%%%%%%%%%%%%%%%%%%%%%%%%%%%%%%%%%%%%%%%%%%%%%%%%%%%%%%%%%%%%%%%%%
, after adjusting the data using a common nuclear suppression parameter
$\alpha=0.96\pm0.01$ ~\cite{e866-alpha} and the latest value of branching ratio
$\mbox{Br}(J/\psi\to l^+l^-)$ \cite{pdg}.
  Eq.\,(\ref{basic}) assumes the same numerical values of $\alpha$ parameters
for $\Upsilon$ and $J/\psi$ production in accordance with the
experimental data~\cite{e866-alpha,alde}.

\section{The HERA-B detector and the data sample}
%=======================

  HERA-B is a fixed-target experiment at the 920-GeV HERA proton storage ring of DESY
and consists of a forward magnetic spectrometer 
featuring a high resolution vertexing and tracking system and offering
a good coverage of the central region of collisions 
(the $x_F$ range is about $[-0.6, 0.15]$ for $\Upsilon$ production and
about $[-0.35, 0.15]$ for $J/\psi$ production,
corresponding to the rapidity ranges about $[-1.2, 0.3]$ 
and $[-1.5, 0.5]$, respectively).

\begin{figure}[tb]
\begin{center}
\epsfxsize=1.05\textwidth\epsfbox{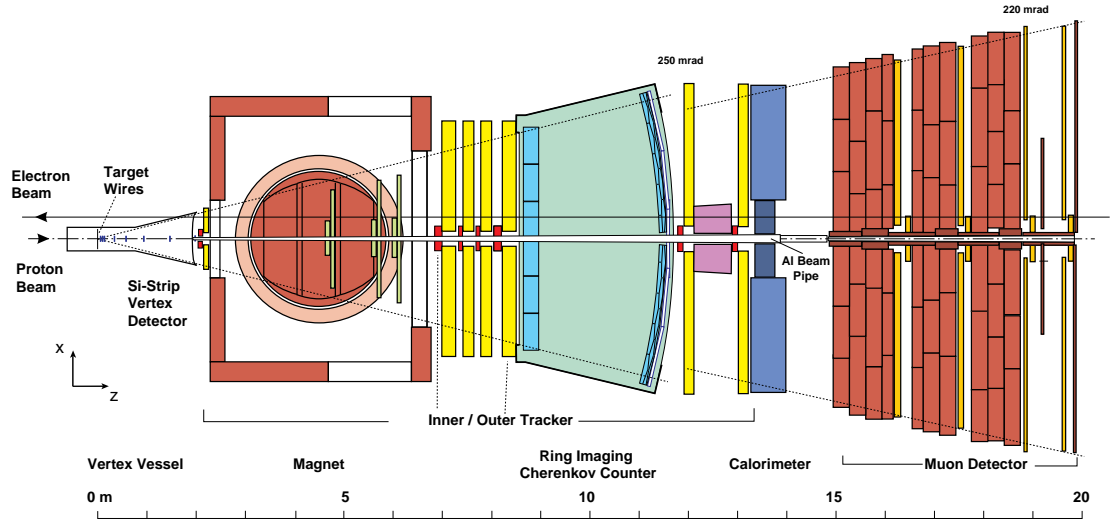}
\caption{Plan view of the HERA-B detector.}
\label{herab-setup}
\end{center}
\end{figure}

  The main components of the detector
are sketched in Fig.\,\ref{herab-setup}.
  The target consists of wires of various materials which are
inserted into the halo of the HERA proton beam.
  Data are taken with carbon, tungsten and titanium target wires operated at an interaction
rate between 5 and 8~MHz. 
  The Vertex Detector System (VDS) consists of silicon micro-strip
detectors located within the vacuum vessel of the target.
  The first station of the main tracker is placed upstream 
of the 2.13~T$\cdot$m 
dipole magnet and
 the remaining 6 tracking stations are placed downstream.
Muon identification is performed by
the muon detector (MUON), while the electrons are
detected and identified by the electromagnetic calorimeter (ECAL).

  The trigger chain includes pretriggers provided by ECAL and MUON
for the lepton candidate search, and a first level trigger (FLT) which
finds tracks downstream of the magnet starting from the pretrigger seeds.
  The FLT requires that at least two pretrigger candidates be present in
an event and that an FLT track be found from at least one of them.
  The (software) second level trigger (SLT) starts from pretrigger candidate tracks,
confirms them in the tracker and VDS using a simplified Kalman filter algorithm, and
accepts the event if either two electron or two muon candidates with a common vertex are found.
The trigger imposes
a cut on the transverse energy $E_T$ of the electrons,
and an implicit cut on the transverse momentum $p_T$ of the muons. 
  A more detailed description of the HERA-B detector, trigger and
reconstruction chain can be found in Refs.\,\cite{herab,bbar-2005}, and references therein.
This analysis is based on
 $134\cdot10^6$  events 
obtained in the 2002--2003 physics run.

%%%%%%%%%%%%%%%%%%%%%%%%%%%%%%%%%%%%%%%%%%%%%%%%%%%%%%%%%%%
%%%%%%%%%%%%%%%%%%%%%%%%%%%%%%%%%%%%%%%%%%%%%%%%%%%%%%%%%%%

%=======================
\section{Data analysis}
%=======================%=======================
\subsection{Monte Carlo simulation, trigger and reconstruction efficiency}
\label{mumu-recoeffi}
%=======================

  Monte Carlo simulation is performed for $J/\psi$, $\Upsilon$ and
Drell-Yan production.
  Heavy quarkonia and Drell-Yan production are simulated with 
PYTHIA 5.7 \cite{PYTHIA}. 
  The energy remaining after the quarkonium or Drell-Yan
generation is passed to FRITIOF 7.02 \cite{FRITIOF}
to simulate the underlying $p$A interaction.
  Finally, additional inelastic $p$A interactions, generated by FRITIOF, are 
superimposed according to the multiple-interaction probabilities
of five separate running periods.
  After tracking the particles through the detector material with GEANT
3.21 \cite{GEANT} and a realistic digitization simulation, the event is
reconstructed by the same reconstruction program which is used for
the real events. 
The trigger and the detector parameters are tuned for the 
individual data taking periods.
To have a realistic description of the kinematic distributions for $J/\psi$
and $\Upsilon$, the PYTHIA generation has been tuned
by reweighting
according to the differential distributions 
from the high-statistics data of Refs.\,\cite{moreno,McGaughey,e789,e771-jpsi}.
The relative trigger and reconstruction efficiencies are determined
from Monte Carlo simulations.  For the muon channel, the ratio of
efficiencies is 
${\varepsilon(J/\psi)/\varepsilon(\Upsilon)} = 0.76\pm0.05$.
  The departure from unity is a result of the trigger
$p_T$ cut which affects muons from $J/\psi$ decay more strongly than
muons from $\Upsilon$ decay due to the different mass scale.  Because
of a harder cut on $E_T$ at the trigger level, the difference is even
larger in the electron channel:
${\varepsilon(J/\psi)/\varepsilon(\Upsilon)} = 0.31\pm0.03$.

\subsection{Event selection}
%=======================

The event selection includes cuts on the quality of tracks,
most notably, that the tracks must have segments in both
the VDS and the main tracker and that ``clones'' 
(nearby reconstructed tracks originating from the same real physical track)
be removed.
Further requirements are that either two muon or two electron candidates of opposite
charge and with a common vertex are present. Muon candidates are required to penetrate to 
the muon detector layers behind the absorber material.
Electrons are identified by requiring an energy deposit (cluster) in the ECAL.
Bremsstrahlung photons emitted by electrons in front of the magnet are recovered
for better energy resolution and
additional electron identification \cite{bbar-2005}.
The transverse energy $E_T$ at ECAL must exceed 0.95~GeV,
and the ratio between the energy as measured
by the ECAL and the momentum must be close to unity 
($0.85<E/p<1.25$ and, in addition, $E/p>0.9$ when no
bremsstrahlung photon is found).

To suppress background from secondaries produced 
in the beam pipe which reach the muon chambers,
we impose cuts on the muon transverse momentum, $p_T$,
which are linear functions of the mass with cut ranges:
$p_T\in[3,6]\mbox{~GeV}/c$ for the invariant mass $m=m_\Upsilon$,
$p_T\in[0.7,4]\mbox{~GeV}/c$ for $m=m_{J/\psi}$
and cuts on the muon total momentum $p$:
$p\in[15,190]\mbox{~GeV}/c$ for $m=m_\Upsilon$,
$p\in[4,100]\mbox{~GeV}/c$ for $m=m_{J/\psi}$;
for intermediate dilepton masses a linear interpolation of the cut limits is used.
 For electrons which do not suffer from a similar background, the cut
for the electron total momentum is $p\in[4,190]\mbox{~GeV}/c$
and for the transverse momentum $p_T\in[0.7,6]\mbox{~GeV}/c$.
After applying these cuts, the lepton momentum spectra 
for the mixture of signal and all background sources 
are similar in the data and Monte Carlo.

%=======================
\subsection{Description of signal and background}
%=======================
\subsubsection{Muon channel}
\label{mumu-channel}
%=======================

%------------------------------------------------------ 
\begin{figure}[t]
\begin{center}
\epsfxsize=0.99\textwidth\epsfbox{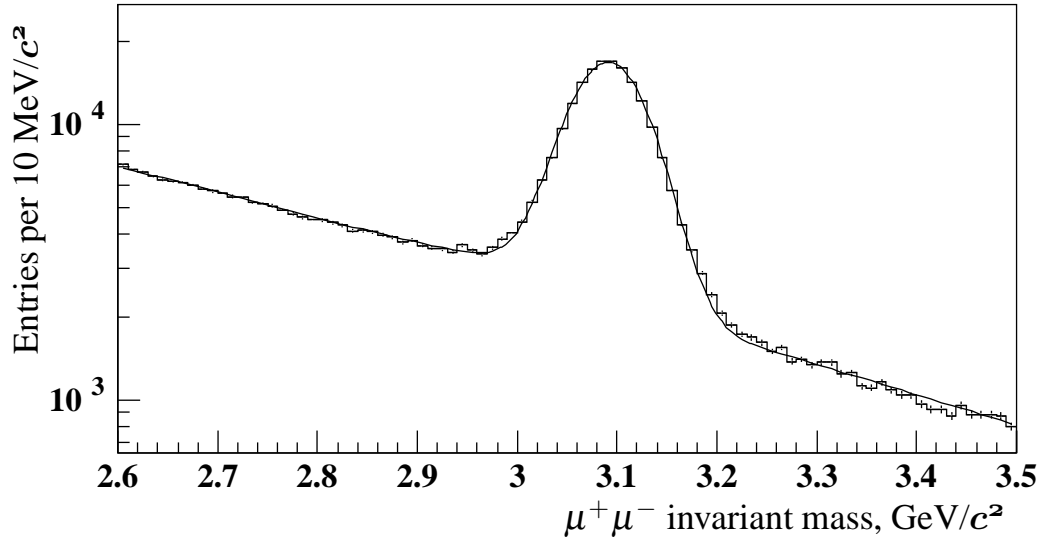}
\caption{Fit of the dimuon mass spectrum obtained after $J/\psi$ selection cuts.}
\label{fig-mumu-jpsi}
\end{center}
\end{figure}
%------------------------------------------------------ 

 The dimuon mass spectrum in the $J/\psi$ mass region (Fig.\,\ref{fig-mumu-jpsi})
is fitted with a Gaussian folded with contributions from 
$J/\psi \rightarrow \mu^+ \mu^- \gamma$~\cite{spiridonovradiative} for
the signal shape, and an exponential
of a second-order polynomial for the background.
For the combined sample of the three target materials
(64\% C, 33\% W, 3\% Ti) 
we obtain a total of $152900\pm490\,J/\psi$ decays 
and a width of 
$38.8\pm0.1$~MeV/$c^2$.

\begin{figure}[hbt]
\begin{center}
\epsfxsize=0.99\textwidth\epsfbox{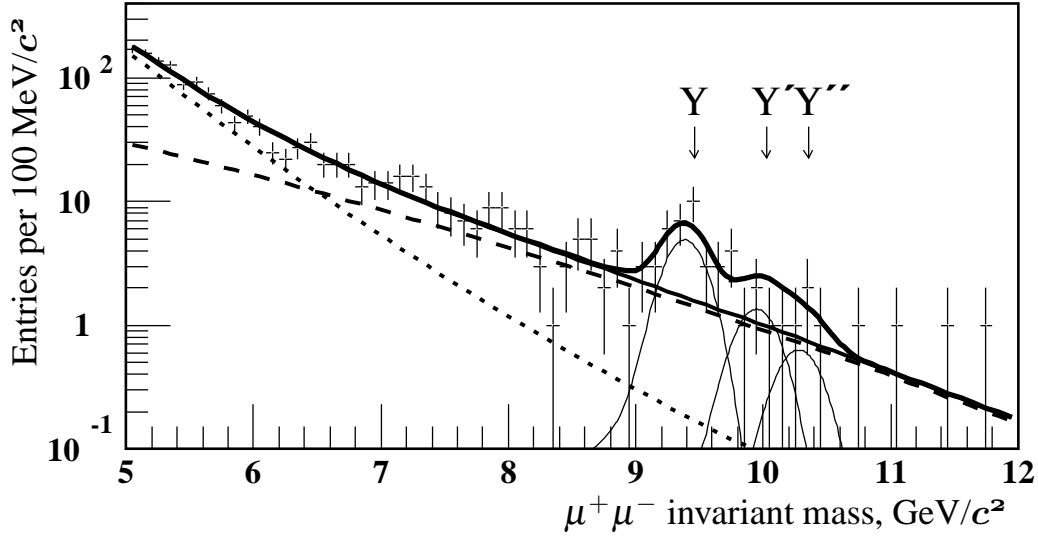}
\caption{Fit of the $\Upsilon\to \mu^+\mu^-$ signal (thick line) 
with the individual contributions of the $\Upsilon(1S)$, $\Upsilon(2S)$, $\Upsilon(3S)$ states
shown in thin solid lines.
  The background consists of Drell-Yan pairs (dashed line) and combinatorial contribution
  estimated from the $\mu^{\pm}\mu^{\pm}$ spectrum (dotted line).}
\label{fig-mumu-fit}
\end{center}
\end{figure}

  For the mass region $m>5$~GeV/$c^2$ (Fig.\,\ref{fig-mumu-fit})
we use a similar function to describe the $\Upsilon$~peaks to that used for the description of the $J/\psi$.
Due to the lack of statistics, the positions of the $\Upsilon(2S)$ and 
$\Upsilon(3S)$ states are fixed relative to the $\Upsilon(1S)$ 
state using the PDG mass values~\cite{pdg}, and
the relative contributions of the three states are fixed according to the 
E605 results~\cite{moreno}:
\begin{equation}
N(1S):N(2S):N(3S)= (70 \pm 3) : (20 \pm 2) : (10 \pm 1).  % 9.8 +- 1.4
\label{e605-fractions}
\end{equation}
  The width of the $\Upsilon(1S)$
state is fixed to the width of the $J/\psi$ scaled by the ratio of
the expected momentum resolution for muons from $\Upsilon$ and $J/\psi$ decays resulting in 159~MeV/$c^2$.
  The widths of the $\Upsilon(2S)$ and $\Upsilon(3S)$ states are scaled
proportionally to their masses.

\begin{figure}[hbt]
\begin{center}
\epsfxsize=0.99\textwidth\epsfbox{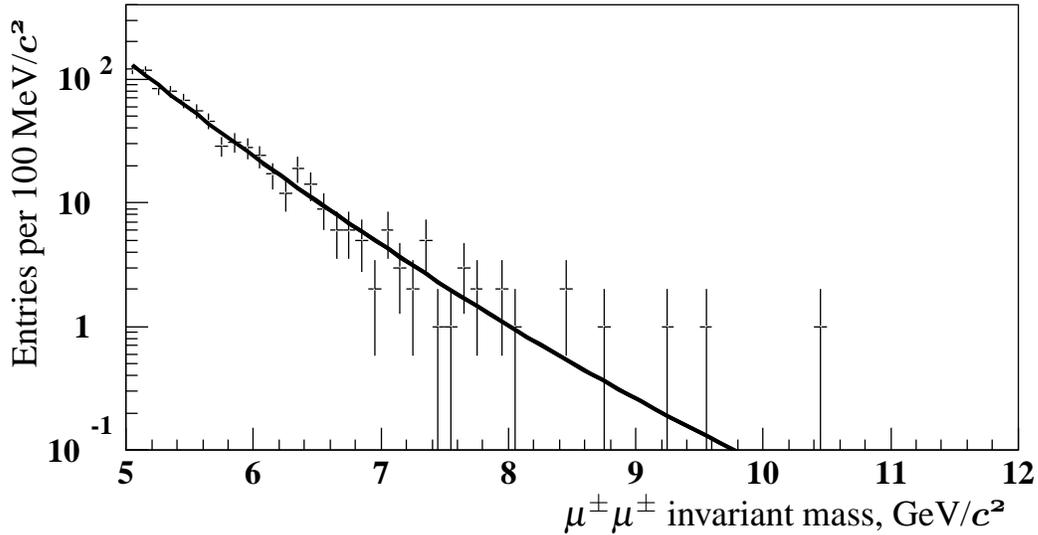}
\caption{Spectrum of like-sign $\mu^\pm\mu^\pm$ pairs, 
together with a fit to an exponential of a second-order polynomial.}
\label{fig-mumu-likesign}
\end{center}
\end{figure}

In this mass region,
the Drell-Yan process as well as random combinations between leptons 
contribute to the background.
 As before, we describe the combinatorial background by
an exponential of a second-order polynomial, whose shape is determined from a 
fit to the like-sign $\mu^\pm\mu^\pm$ pair spectrum~(Fig.\,\ref{fig-mumu-likesign}),
and a normalization factor which is left as a free parameter.
  The shape of the Drell-Yan spectrum is determined from a fit to the
corresponding reconstructed Monte Carlo events.

%------------------------------------------------------ 
%------------------------------------------------------ 
%------------------------------------------------------ 
%------------------------------------------------------

Thus, the free fit parameters for the spectrum 
with $m>5$~GeV/$c^2$ (Fig.\,\ref{fig-mumu-fit}) are:
the total $\Upsilon$~yield, the $\Upsilon(1S)$ mass,
the yield of Drell-Yan dileptons, and the height of the combinatorial background.
The total $\Upsilon$~yield in the log likelihood fit is
$N(\Upsilon) = 30.8 \pm 7.4 _{\rm stat}$.
The fit function describes the data well ($\chi^2/\mbox{ndf}=51/54$).
Using this value in Eq.\,(\ref{basic}) we obtain
$R^{\mu\mu}_{J/\psi} = (7.9 \pm 1.9_{\rm stat})\cdot10^{-6}$,
and using the value of the $J/\psi$ cross section Eq.\,(\ref{jpsi-reference}),
the cross section for $\Upsilon$~production at mid-rapidity as determined from 
$\Upsilon \rightarrow \mu^+\mu^-$
becomes 
$\mbox{Br}(\Upsilon\to\mu^+\mu^-) \cdot {d\sigma(\Upsilon)/dy}|_{y=0} =  4.0 \pm 1.0_{\rm stat}\,{\mbox{pb}/\mbox{nucleon}}$.

  The fit (Fig.\,\ref{fig-mumu-fit}) shows that the combinatorial background dominates below
7~GeV/$c^2$ but becomes negligible at the $\Upsilon$~mass.
  We have verified that the generated shape of the Drell-Yan background 
in Monte Carlo simulations agrees with the published data
\cite{moreno,McGaughey,nusea-drell-yan}.
  Further support for the correctness of the background curves in
Fig.\,\ref{fig-mumu-fit} comes from studying 
a number of sensitive kinematic variables 
(such as the average momentum asymmetry between the muons).
The observed values
are compatible with the mixture of signal and background contributions
obtained from the fit to the mass spectrum.

The fit
 also provides a measurement of the Drell-Yan (DY)
differential cross section $d^2\sigma/(dm\,dy)|_{y=0}$ in the region
of the $\Upsilon$~mass, or equivalently
the ratio $R_{\rm DY}$~\cite{childress,moreno,yoshida} of
the $\Upsilon$~production to the Drell-Yan cross sections
$$\displaystyle R_{\rm DY}=
{\mbox{Br}(\Upsilon\to l^+l^-)\cdot{d\sigma\over dy}(\Upsilon)\Big|_{y=0} \,\,\Big/\,\,
 \displaystyle {d^2\sigma \over dm\,dy}({\rm DY}){\Big|}_{\rput(2.2ex,3.5ex){\mbox{\scriptsize $y=0$}}m=9.46}}\;\;.$$
Similarly to Eq.\,(\ref{basic})
we determine for the Drell-Yan cross section: 
${ {d^2\sigma \over dm\,dy}({\rm DY})\big|
  _{\rput(2.2ex,2.5ex){\mbox{\scriptsize $y=0$}}m=9.46}}=
2.0 \pm 0.5 \,{\mbox{pb}/(\mbox{nucleon}\cdot\mbox{GeV/$c^2$}})$
and $R^{\mu\mu}_{\rm DY} = 2.0 \pm 0.8 \mbox{~GeV/$c^2$}$
where statistical and systematic uncertainties have been combined.

  To compare these results with the published data of the Fermilab
fixed-target experiments \cite{moreno,yoshida,McGaughey,nusea-drell-yan}
at $\sqrt{s}=38.8$~GeV, we scale the Drell-Yan
production cross section with the variable $\tau=m^2/s$ leading to
a factor of $1.28\pm0.03$%
\footnote{This number agrees well with the theoretical
  prediction~\cite{vogt-private} of about 1.275 for
the mass region $8<m<11$~GeV/$c^2$ .},
and the $\Upsilon$~cross section by a factor 1.32
(obtained from the fit to the $\Upsilon$~data, see Sect.\,\ref{sect-fits}). 
The results, scaled to $\sqrt{s}=38.8$~GeV, then become:
${{d^2\sigma \over dm\,dy}({\rm DY})\Big|
       _{\rput(2.2ex,1.9ex){\mbox{\scriptsize $y=0$}}m=9.46}
       ^{\sqrt{s}=38.8}}=
1.5 \pm 0.4 \,{\mbox{pb}/(\mbox{nucleon}\cdot\mbox{GeV/$c^2$}})$
and $R^{\mu\mu}_{\rm DY}|^{\sqrt{s}=38.8} = 2.0  \pm 0.8  \mbox{~GeV/$c^2$}$,
  which agrees within the uncertainties with the fits of the data from
the Fermilab experiments:
$R_{\rm DY}$ between 1.3 and 1.6~GeV/$c^2$
\cite{moreno,yoshida}
and 
${{d^2\sigma \over dm\,dy}({\rm DY})\big|
  _{\rput(2.2ex,2.5ex){\mbox{\scriptsize $y=0$}}m=9.46}}=
1.4 \pm 0.3 \,{\mbox{pb}/(\mbox{nucleon}\cdot\mbox{GeV/$c^2$}})$
\cite{moreno,McGaughey,nusea-drell-yan}.

%=======================
\subsubsection{Electron channel}
%=======================

 The analysis of dielectron events generally follows the path
described for dimuon events.
  Before fitting, the momentum vectors of the leptons are corrected by
adding in the energy of bremsstrahlung photons emitted in the material
before the magnet and reconstructed in the calorimeter.
  The fit function takes into account resolution effects as well as a
tail due to non-recovered bremsstrahlung and final state radiation \cite{bbar-2005,spiridonovradiative}.
The fit yields $N(J/\psi)=109710\pm930$
and a width of 
$59.9\pm0.8\mbox{~MeV/$c^2$}$
for the combined sample of the three target materials
(62\% C, 32\% W, and 6\% Ti).

%------------------------------------------------------ 
\begin{figure}[hbt]
\begin{center}
\epsfxsize=0.99\textwidth\epsfbox{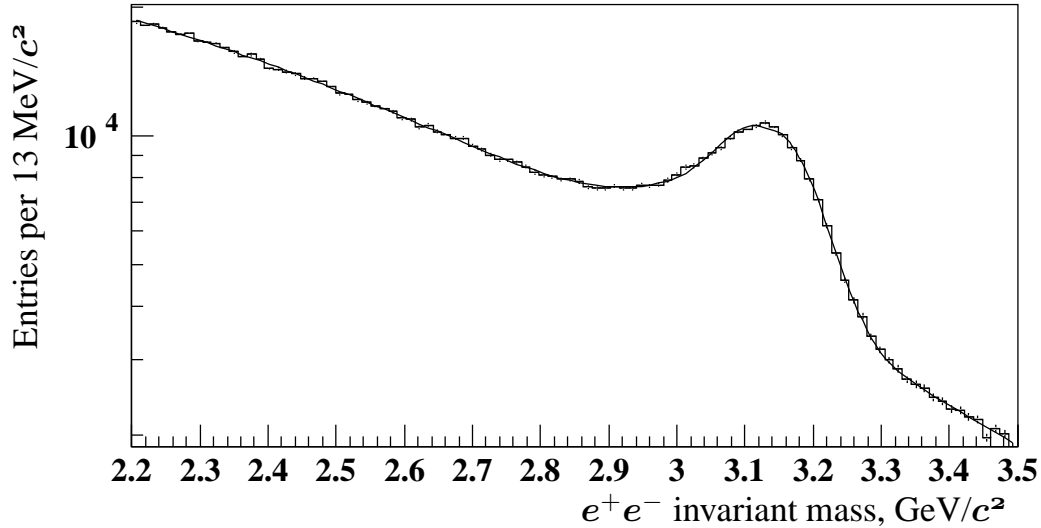}
\caption{Fit of the dielectron mass spectrum obtained after $J/\psi$ selection cuts.}
\label{fig-ee-jpsi}
\end{center}
\end{figure}
%------------------------------------------------------ 
%------------------------------------------------------ 
\begin{figure}[hbt]
\begin{center}
\epsfxsize=0.99\textwidth\epsfbox{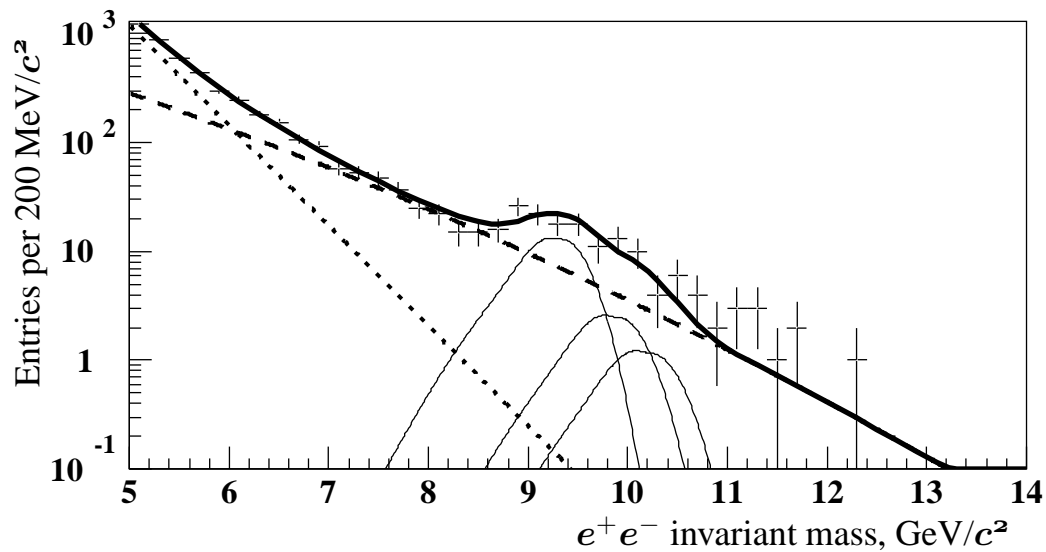}
\caption{Fit of the $\Upsilon\to e^+e^-$ signal (thick line) 
with the individual contributions of the $\Upsilon(1S)$, $\Upsilon(2S)$, $\Upsilon(3S)$ states
shown in thin solid lines.
  The background consists of Drell-Yan (dashed line) and combinatorial contributions (dotted line).
}
\label{fig-ee-fit}
\end{center}
\end{figure}
%------------------------------------------------------ 

The log likelihood fit of the
mass region $m>5$~GeV/$c^2$
(Fig.\,\ref{fig-ee-fit}), for which the free parameters are
the position of the $\Upsilon(1S)$ peak, the contributions of the $\Upsilon$~signal,
of Drell-Yan, and of the combinatorial background,
results in $75 \pm 14 _{\rm stat}$ $\Upsilon$~events with $\chi^2/\,\mbox{ndf}=19/27$.
This leads to a cross section ratio
$R^{ee}_{J/\psi}=(11.0 \pm 2.1_{\rm stat})\cdot 10^{-6}$,
and a mid-rapidity cross section for $\Upsilon$~production of
$\mbox{Br}(\Upsilon\to e^+ e^-)\cdot {d\sigma(\Upsilon)/dy}\big|_{y=0} = 
5.5 \pm 1.0_{\rm stat}\,\, {\mbox{pb}/\mbox{nucleon}}$.

  As for the muon channel, the various energy-dependent kinematic
distributions of the electron pairs behave as expected from the
relative contributions of the different processes.
  The imposition of an $E_T$ requirement for electrons in the trigger
severely suppresses triggers from like-sign leptons, because of the
additional $p_T$-bias of the magnet.
The resulting like-sign electron mass spectrum therefore does not
describe the combinatorial background and is not used in this analysis.
Instead, the combinatorial background is evaluated by mixing
tracks from different events.
For masses greater than about 6~GeV/$c^2$, the combinatorial background is
less than the Drell-Yan contribution and becomes negligible for masses
above 9~GeV/$c^2$.

The Drell-Yan contribution at $m=9.46$~GeV/$c^2$ is
${{d^2\sigma \over dm\,dy}({\rm DY})\Big|
   _{\rput(2.2ex,2.5ex){\mbox{\scriptsize $y=0$}}m=9.46}} =
3.7 \pm 1.0\,{\mbox{pb}/(\mbox{nucleon}\cdot\mbox{GeV/$c^2$}})$
leading to $R^{ee}_{\rm DY}=1.5\pm0.7\mbox{~GeV/$c^2$}$.
Scaling these results to $\sqrt{s}=38.8$~GeV we obtain
${{d^2\sigma \over dm\,dy}({\rm DY})
\Big|_{\rput(2.2ex,2.5ex){\mbox{\scriptsize $y=0$}}m=9.46}
%   ^{\sqrt{s}=38.8}
}= 2.9 \pm 0.8 \,{\mbox{pb}/(\mbox{nucleon}\cdot\mbox{GeV/$c^2$})}$,
and
$R^{ee}_{\rm DY}|^{\sqrt{s}=38.8} = 1.4 \pm 0.6 \mbox{~GeV/$c^2$}$.
  The Drell-Yan cross section measurements for the electron and muon cases differ by a
factor of 1.9, corresponding to 1.6~standard deviations;
the ratio $R_{DY}^{ee}$ agrees with the muon case and with Fermilab measurements.

%=======================
\subsection{Systematic uncertainties}
\label{sect-syst}
%=======================

The systematic uncertainties 
are dominated by the uncertainties in the description of the background
which contribute 14\% for muons
and 17\% (including uncertainties in the bremsstrahlung tail) for electrons, respectively.
The uncertainties of the relative $J/\psi$ and $\Upsilon$~efficiencies are estimated to be 
7\% for muons and 9\% for electrons.
The systematic uncertainty of the $J/\psi$ reference cross section Eq.\,(\ref{basic}) is 9\%. 
  The parameter $\Delta y_{\rm eff}(\sqrt{s}) =1.14 \pm 0.12_{\rm syst}$ (precision of 11\%)
at $\sqrt{s} = 41.6$~GeV is determined from the fits of existing 
measurements for $x_F$ and $p_T$ distributions for $\Upsilon$ mesons \cite{moreno,McGaughey}.

Other systematic uncertainties are small compared with
those already mentioned. Among them are
the use of a Gaussian function with a fixed width 
for fitting of the $\Upsilon$ peaks
($<4\%$),
uncertainties of the fractions of the various $\Upsilon$~states in Eq.(\ref{e605-fractions})
($<0.1\%$),
the polarization effects in $J/\psi$ and $\Upsilon$ production ($<1.8\%$),
and the branching ratio $J/\psi \to l^+l^-$
($<1.7\%$).
When, in the electron channel, the particle identification
requirements are strengthened by requiring either that one of the two
lepton candidates have an associated bremsstrahlung cluster in the
calorimeter, or that both have associated clusters, the results change
by less than 4\%.
 The results are stable 
within the statistical uncertainty 
for a
wide variation of the cuts for muon or electron identification.
All contributions, added in quadrature, result in a systematic uncertainty 
of 21\% in the case of muons, and 25\% in the case of electrons.

%==============================================================================
\begin{table*}[hbt]
\caption{Summary of the input values and the resulting cross sections 
         for both muon and electron channels.}
\label{tab-firstsummary}
\scriptsize
\small
\scalebox{1.0}{
\begin{tabular}{|c|c|c|}
\hline
Parameter            & $\mu^+\mu^-$ channel & $e^+e^-$ channel \\ \hline \hline
$N(\Upsilon)$    & $30.8 \pm 7.4$        & $75 \pm 14$   \\ \hline
$N(J/\psi)$      & $ 152900 \pm 490$ & $ 109710 \pm 930$ \\ \hline
$\varepsilon(J/\psi)/\varepsilon(\Upsilon)$ & $0.76 \pm 0.05$  & $0.31 \pm 0.03$ \\ \hline 
\hline
$\sigma_{pN}(J/\psi)$ & \multicolumn{2}{|c|}{$502\pm 44$}\\[-1.2ex] 
 (nb/nucleon)                                                          &  \multicolumn{2}{|c|}{}  \\ \hline
$\mbox{Br}(J/\psi\to l^+l^-)$   & \multicolumn{2}{|c|}{$(5.90\pm0.10)\%$} \\ \hline
$\Delta y_{\rm eff}$ & \multicolumn{2}{|c|}{$1.14 \pm 0.12_{\rm syst}$} \\ \hline
\hline
$R_{J/\psi}$ (units of $10^{-6}$)    & $7.9 \pm 1.9_{\rm stat} \pm 1.5_{\rm syst}$  & 
                                               $11.0\pm 2.1_{\rm stat} \pm 2.5_{\rm syst}$ \\ \hline
\raisebox{-1ex}{$\displaystyle\mbox{Br}(\Upsilon\to{}l^+l^-)\cdot{d\sigma\over dy}(\Upsilon)\,\big|_{y=0}$}
                                     & $4.0 \pm 1.0_{\rm stat} \pm 0.8_{\rm syst}$ & $5.5 \pm 1.0_{\rm stat} \pm 1.4_{\rm syst}$   \\[-1.5ex]
(pb/nucleon)  & & \\ \hline
 \raisebox{-1.7ex}{$\displaystyle {d^2\sigma({\rm DY})\over dm\,dy}$}
      &   $2.0 \pm 0.5$ &  $3.7 \pm 1.0 $ \\[-1.4ex]
($\mbox{pb}/\mbox{nucleon}\cdot \mbox{GeV}/c^2) $  && \\ \hline
$R_{\rm DY}$  &   $2.0  \pm 0.8 $ &  $1.5  \pm 0.7 $ \\[-0.6ex]
 & & \\ \hline
\end{tabular}}
\end{table*}
%=======================

\section{Combined results}
\label{sect-fits}
%==================

\begin{figure}[t]
\begin{center}
\epsfxsize=0.99\textwidth\epsfbox{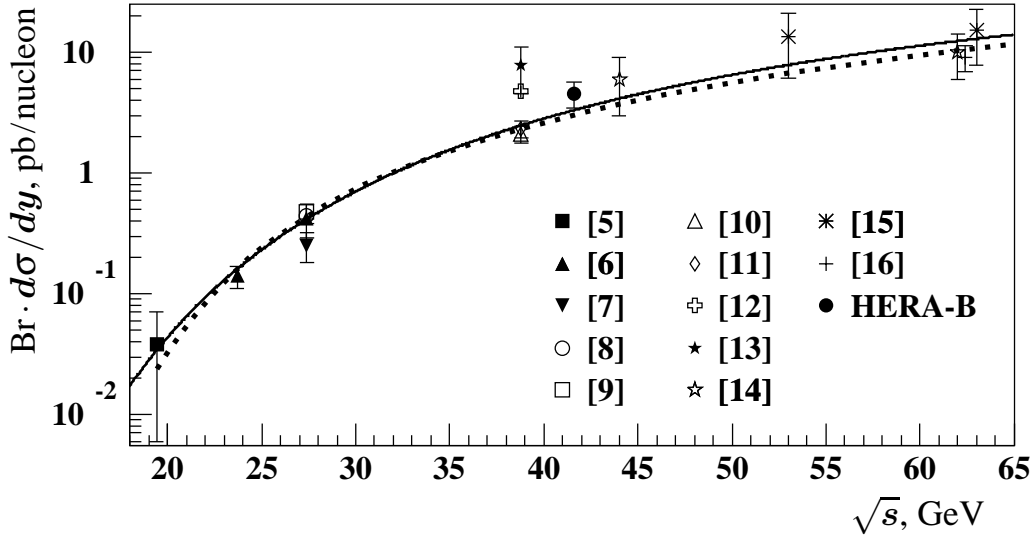}
\caption{This result (HERA-B) compared to the fit (solid curve) of the data of previous experiments 
\cite{badier}--\cite{angelis} up to $\sqrt{s}=63$~GeV.
The NLO theoretical curves (see description in the text), shown 
as a dashed line, differ by less than the width of the line.
}
\label{fig-final}
\label{upsfit-all}
\end{center}
\end{figure}

Table \ref{tab-firstsummary} summarizes all results obtained
in the previous sections.
One can see that the results in the muon and electron channels are compatible.
Combining the muon and electron channels,
we have
$$
{R_{J/\psi} = (9.0 \pm 2.1) \cdot 10^{-6}}
$$
and, using the $J/\psi$ reference cross section Eq.\,(\ref{jpsi-reference}),
we obtain (Fig.\,\ref{fig-final})
\begin{equation}
\mbox{Br}(\Upsilon\to l^+l^-)\cdot{d\sigma\over dy}(\Upsilon)\,\Big|_{y=0}= 4.5 \pm 1.1 \,\,{\mbox{pb}\over\mbox{nucleon}}\,\,
\label{res-syst-combined}
\end{equation}
where the error includes both statistical and systematic contributions.
The systematic
uncertainties in the muon and electron channel have a common part
of 14\% compared to the full values of 21\% and 25\%, respectively,
the dominant contributions being due to the uncertainties
of $\Delta y_{\rm eff}$ and the $J/\psi$ reference
cross section Eq.\,(\ref{jpsi-reference}).
After taking this into account,
the $\chi^2$ of the combined result is $\chi^2=0.6$.

%%%%%%%%%%%%%%%%%%%%%%%%%%%%%%%%%%%%%%%%%%%%%%%%%%%%%%%%%%%%%%%%%%%%%
%%%%%%%%%%%%%%%%%%%%%%%%%%%%%%%%%%%%%%%%%%%%%%%%%%%%%%%%%%%%%%%%%%%%%

If, instead of normalizing to the $J/\psi$ cross section,
we use the yield and efficiency ratio
between $\Upsilon$ and Drell-Yan production
together with the DY cross section at $\sqrt{s}=38.8$~GeV of 
${d^2\sigma \over dy\,dm}({\rm DY})\big|_{y=0} = (1.35 \pm 0.27)
\,\,{{\rm pb}/({\rm nucleon}\cdot{\rm GeV}/c^2)}$~\cite{moreno}, scaled by 
the factor $1.28 \pm 0.03$ to $\sqrt{s}=41.6$~GeV, we obtain for the
 $\Upsilon$ production cross section:
${\mbox{Br}(\Upsilon\to{}l^+l^-) \cdot {d\sigma\over dy}(\Upsilon)\Big|_{y=0}} =
 2.9 \pm 1.1\,\,{{\rm pb}\over{\rm nucleon}}$
with a $\chi^2$ of 0.3.
However, 
the published data on 
the DY process are less abundant than those for the $J/\psi$ cross section and 
are less precise due to limited
statistics and difficulties in separating the $\Upsilon$ signal from the DY continuum.

We therefore take the value obtained by normalization to the DY cross
section as a confirmation of our main result Eq.\,(\ref{res-syst-combined}), which is
obtained by normalization to the $J/\psi$ cross section.
  Scaling this result by a factor of 1.32
(obtained from the fit described below)
 to compare it to the published data at 
$\sqrt{s}=38.8$~GeV, we obtain:
$\mbox{Br}(\Upsilon\to l^+l^-)\cdot{d\sigma\over dy}(\Upsilon)\big|_{y=0}^{\sqrt{s}=38.8}
   = 3.4 \pm 0.8 \,\,{\mbox{pb}/\mbox{nucleon}}.$
  Our result lies half-way between the results of E605 \cite{moreno,yoshida}
and E772 \cite{McGaughey}
and does not
favour one of these results over the other.
  In Fig.\,\ref{fig-final} we compare the result Eq.\,(\ref{res-syst-combined}) 
with the measurements of previous experiments. 
  In most cases, the uncertainties are large with the
exception of four measurements obtained at Fermilab at cms energy $\sqrt{s}=38.8$~GeV, 
which are also those closest in energy to \mbox{HERA-B} ($\sqrt{s}=41.6$~GeV).
  However, these values are in poor agreement with each other,
as summarized in Table\,\ref{tab-exper}.

  The fit of all experiments in Fig.\,\ref{upsfit-all} (solid line) uses the Craigie parameterization \cite{craigie}
$f(\sqrt{s})=\sigma_0 \exp\left(-{m_0/\sqrt{s}}\right)$,
yielding 
$\sigma_0 = 182  \pm 21 $~pb/nucleon,
$m_0      = 167  \pm  4 $~GeV,
$\chi^2/\mbox{ndf}=37/14$.
 The dotted line in Fig.\,\ref{upsfit-all} shows fits 
to predictions of next-to-leading order (NLO) calculations 
from Ref.\,\cite{vogt-hard-probes}
in the framework of the colour evaporation model (CEM)  using the MRST~HO parton
distribution functions \cite{mrst}.
 The thickness of the line corresponds to the 
variations of results in the three sets of input parameters of the model
which describe the open beauty production data \cite{vogt-hard-probes}:
$m_b=\mu=4.75$~GeV/$c^2$; $m_b=4.5$~GeV/$c^2$, $\mu=2m_b$; and $m_b=5$~GeV/$c^2$, $\mu=m_b/2$,
where $m_b$ is the $b$-quark mass and $\mu$ is 
the renormalization scale
(assumed to be equal to the factorization scale).
The NLO predictions are
normalized according to a fit of the experimental data 
using the mentioned values of the parameters
$m_b$ and $\mu$ \cite{vogt-hard-probes}.

%=======================
\section{Conclusion}
%=======================

  The $\Upsilon$ production yield at
mid-rapidity in $p$A collisions at a
proton momentum $p=920$~GeV/$c$ has been measured
in both channels  $\Upsilon\to\mu^+\mu^-$ and $\Upsilon\to{}e^+e^-$.
The $J/\psi$ cross section (\ref{jpsi-reference}) has been used for normalization,
and the ratio of $\Upsilon$ and $J/\psi$ cross sections Eq.\,(\ref{eq-rjpsi}) is determined to be
$$
R_{J/\psi}=(9.0 \pm 2.1)\cdot 10^{-6}.
$$
The resulting $\Upsilon$ production cross section (both
lepton channels combined) is
$$
\mbox{Br}(\Upsilon\to l^+l^-)\cdot{d\sigma\over dy}(\Upsilon)\,\Big|_{y=0}= 
4.5 \pm 1.1 \,\,{\mbox{pb}\over\mbox{nucleon}}\,\,.
$$
Our result, scaled for $\sqrt{s}$ dependence, 
lies half-way between those of E605 \cite{moreno,yoshida}
and E772 \cite{McGaughey}.
 Normalization with respect to the Drell-Yan process confirms this result.
The result agrees within 1.4 standard deviations
with current CEM NLO predictions \cite{vogt-hard-probes} (see Fig.\,\ref{fig-final}).

%------------------------------------------------------ 
\vspace{5mm} 
{\bf Acknowledgments:}

  We express our gratitude to the DESY laboratory and to the DESY
accelerator group for their strong support since the conception of the
HERA-B experiment.
  The HERA-B experiment would not have been possible without the enormous
effort and commitment of our technical and administrative staff. It is a
pleasure to thank all of them.
  We would like to thank R.\,Vogt for many stimulating discussions and suggestions.

 \def\largelinestretch{\renewcommand{\baselinestretch}{0.93}}
 \largelinestretch\small\normalsize
%====================================================
%
%                       References
%
%====================================================

\end{document}

%% file: authors_epjc.tex
I.~Abt$^{23}$,
M.~Adams$^{10}$,
M.~Agari$^{13}$,
H.~Albrecht$^{12}$,
A.~Aleksandrov$^{29}$,
V.~Amaral$^{8}$,
A.~Amorim$^{8}$,
S.~J.~Aplin$^{12}$,
V.~Aushev$^{16}$,
Y.~Bagaturia$^{12,36}$
% ...

\vspace{5mm}
\noindent
% ...
$^{8}${\it LIP Coimbra, P-3004-516 Coimbra,  Portugal~$^{f}$} \\
% ...
$^{10}${\it Institut f\"ur Physik, Universit\"at Dortmund, D-44221 Dortmund, Germany~$^{d}$} \\
% ...
$^{12}${\it DESY, D-22603 Hamburg, Germany} \\
$^{13}${\it Max-Planck-Institut f\"ur Kernphysik, D-69117 Heidelberg, Germany~$^{d}$} \\
% ...
$^{16}${\it Institute for Nuclear Research, Ukrainian Academy of Science, 03680 Kiev, Ukraine~$^{h}$}\hspace*{-18mm}~ \\
% ...
$^{23}${\it Max-Planck-Institut f\"ur Physik, Werner-Heisenberg-Institut, D-80805 M\"unchen, Germany~$^{d}$} \\
% ...
$^{29}${\it DESY, D-15738 Zeuthen, Germany} \\
% ...
$^{36}${\it visitor from High Energy Physics Institute, 380086 Tbilisi, Georgia} \\
% ...
$^{d}$ supported by the Bundesministerium f\"ur Bildung und Forschung, FRG, under contract numbers 05-7BU35I, 05-7DO55P, 05-HB1HRA, 05-HB1KHA, 05-HB1PEA, 05-HB1PSA, 05-HB1VHA, 05-HB9HRA, 05-7HD15I, 05-7MP25I, 05-7SI75I \\
% ...
$^{f}$ supported by the Portuguese Funda\c c\~ao para a Ci\^encia e Tecnologia under the program POCTI \\
% ...
$^{h}$ supported by the National Academy of Science and the Ministry of Education and Science of Ukraine \\
% ...


%% file: ups.bbl
\begin{thebibliography}{99}
%\vspace{-2mm}
%\newcommand{\tit}[1]{{}}
\addcontentsline{toc}{section}{\mbox{References}}

\bibitem{vogt} R. Vogt, Phys.~Rept. 310 (1999) 197.

\bibitem{QWG}
N. Brambilla et al., CERN Yellow Report CERN-2005-005, hep-ph/0412158.
%\tit{``Heavy quarkonium physics''.}

\bibitem{QGP1}
J.F. Gunion and R. Vogt, Nucl. Phys. B492 (1997) 301.
%\tit{``Determining the existence and nature of the quark-gluon plasma by $\Upsilon$ suppression at the LHC.''}

\bibitem{QGP2}
F. Karsch, M.T. Mehr and H. Satz, Z. Phys. C37 (1988) 617.
%\tit{``Color screening and deconfinement for bound states of heavy quarks.''}

%%%%%%%%%%%%%%%%%%%%%%%%%%%%%%%%%%%%%%%%%%%%%%%%%%%%%%%%%%%%%%%% Experiments
\newcommand{\collaboratio}[1]{#1} % To restore collaboration names in the references

%\tit{\\ A high-statistics study of $\Upsilon$-meson production in pi-W reactions at 286 GeV/$c$.}
\bibitem{badier} \collaboratio{NA3 collab.,} J. Badier et al., Phys.~Lett. B86 (1979) 98.
%\tit{\\ First evidence for $\Upsilon$ production by pions.}
\bibitem{yoh} J.K. Yoh et al., Phys.~Rev.~Lett. 41 (1978) 684; Erratum ibid. 41 (1978) 1083. 
\bibitem{innes} W.R. Innes et al., Phys.~Rev.~Lett. 39 (1977) 1240; 
 Erratum ibid. 39 (1977) 1640. 
\bibitem{ueno} K. Ueno et al., Phys.~Rev.~Lett. 42 (1979) 486.
\bibitem{childress} S. Childress et al., Phys.~Rev.~Lett. 55 (1985) 1962.
%\tit{\\ Production dynamics of the $\Upsilon$ in proton-nucleon interactions.}

\bibitem{moreno} \collaboratio{E605 collab.,} G. Moreno et al., Phys.~Rev. D43 (1991) 2815.
%\tit{\\ dimuon production in proton - copper collisions at $s^{(1/2)} = 38.8$ GeV.}

\bibitem{yoshida} \collaboratio{E605 collab.,} T. Yoshida et al., Phys.~Rev. D39 (1989) 3516.
%\tit{\\ High resolution measurement of massive dielectron production in 800 GeV proton beryllium collisions.}

\bibitem{McGaughey}
\collaboratio{E772 collab.,} P.L. McGaughey et al., Phys.~Rev. D50 (1994) 3038; Erratum ibid. D60 (1999) 119903.
%\tit{\\ cross-sections for the production of high mass muon pairs from 800 GeV proton bombardment of h-2.}

\bibitem{e771} \collaboratio{E771 collab.,} T. Alexopoulos et al., Phys.~Lett. B374 (1996) 271.

\bibitem{camilleri} \collaboratio{R-108 (CCOR collab.),} L. Camilleri,
        in: T.B.W. Kirk, H.D.I. Abarbanel (Eds.),
         Proc. 1979 Int. Symp. on Lepton and Photon Interactions at High Energies,
         \mbox{Fermilab}
(1979), p.282. 
\bibitem{kourkoumelis} C. Kourkoumelis et al., Phys.~Lett. B91 (1980) 481. 
%\tit{\\ Characteristics of J/$\psi$ and $\Upsilon$ production at the CERN intersecting storage rings.}
\bibitem{angelis} CCOR collab., A.L.S. Angelis et al., Phys.~Lett. B87 (1979) 398. 
%\tit{\\ A measurement of the production of massive e+ e- pairs in proton proton collisions at s**(1/2) = 62.4 GeV.}
\bibitem{albajar} \collaboratio{UA1 collab.,} C. Albajar et al., Phys.~Lett. B186 (1987) 237. 
%\tit{\\ Beauty production at the CERN proton-antiproton collider.}

\bibitem{cdf} \collaboratio{CDF collab.,} F.~Abe et al., Phys.~Rev.~Lett. 75 (1995) 4358.

%%%%%%%%%%%%%%%%%%%%%%%%%%%%%%%%%%%%%%%%%%%%%%%%%%%%%%%%%%%%%%%%%%%%%%%%%%%%%%%%%%%%%%%%%%

\bibitem{e866-alpha} \collaboratio{E866/NuSea collab.,} M.J. Leitch et al., Phys.~Rev.~Lett. 84 (2000) 3256.

\bibitem{pdg} S.~Eidelman et al., Phys.~Lett. {B592} (2004) 1.
  
\bibitem{jpsi-reference} 
F. Maltoni et al.,
%{Analysis of Charmonium Production at Fixed-Target Experiments in the NRQCD Approach},
% Preprint DESY,
 hep-ph/0601203 (2003), submitted to Phys.~Lett.~B.

\bibitem{herab-jpsi} 
\collaboratio{HERA-B collab.,} I. Abt et al., 
% {``Measurement of the $J/\psi$ Production Cross Section in 920~GeV$/c$ Fixed-Target Proton-Nucleus Interactions''},
Preprint DESY 05-232, Hamburg (2005), hep-ex/0512029, submitted to Phys.~Lett.~B.

\bibitem{alde}
E772 collab., D.M. Alde et al., Phys.~Rev.~Lett. 64 (1990) 2479.

\bibitem{herab} \collaboratio{HERA-B collab.,} E.~Hartouni {et al.},
  HERA-B Design Report, DESY-PRC 95-01 (1995);
% can be found at\\{http://www-hera-b.desy.de/general/publications/proposal/tdr/tdr.ps.gz}
 HERA-B collab., HERA-B Status Report, DESY-PRC 00-04 (2000).
% \\can be found at {http://www-hera-b.desy.de/general/publications/hb2k/}

\bibitem{bbar-2005} 
\collaboratio{HERA-B collab.,} I. Abt et al., 
% {``Improved measurement of the $b\bar{b}$ Production Cross Section in 920~GeV Fixed-Target Proton-Nucleus Collisions''}, 
Preprint DESY 05-233, Hamburg (2005), hep-ex/0512030, Phys.~Rev.~D (in press).

\bibitem{PYTHIA} T.~Sj\"ostrand, Comp.~Phys.~Comm. {82} (1994) 74.
\bibitem{FRITIOF} H.~Pi, Comp.~Phys.~Comm. {71} (1992) 173.
\bibitem{GEANT}  R.~Brun {et al.}, GEANT3, CERN-DD-EE-84-1 (1987).

\bibitem{e789} \collaboratio{E789 collab.,} M.H.~Schub et al., Phys.~Rev. D52 (1995) 1307;
Erratum ibid. D53 (1996) 570.
%\tit{\\ Measurement of J/$\psi$ and $\psi$-prime production in 800 GeV/$c$ proton - gold collisions.}
\bibitem{e771-jpsi} \collaboratio{E771 collab.,} T. Alexopoulos et al., Phys.~Rev. D55 (1997) 3927.

\bibitem{spiridonovradiative} A. Spiridonov, 
% ``Bremsstrahlung in Leptonic Onia Decays: Effects on Mass Spectra'',
Preprint DESY 04-105, Hamburg (2004), hep-ex/0510076.

\bibitem{nusea-drell-yan} 
\collaboratio{E866/NuSea collab.,} J.C. Webb et al., 
%Fermilab-Pub-03-302-E (2003);
hep-ex/0302019 (2003);
%\tit{\\Submitted to Phys. Rev. Lett. }
%\tit{\\Absolute Drell-Yan dimuon cross-sections in 800 GeV/$c$ pp and pd collisions.}
\\
J.C. Webb, 
% FERMILAB-THESIS-2002-56 (2003);
hep-ex/0301031 (2003);
%\tit{\\ Measurement of continuum dimuon production in 800 GeV/$c$ proton-nucleon collisions.}
\\
E866/NuSea Collab., C.A. Gagliardi et al.,
AIP Conf.~Proc. 698 (2004) 100.
% Measurement of the Absolute Drell-Yan Dimuon Cross Sections in 800-GeV/c p p and p d collisions.
% Prepared for 16th International Conference on Particles and Nuclei (PANIC 02), Osaka, Japan, 30 Sep - 4 Oct 2002.

\bibitem{vogt-private} R.~Vogt, private communication (Feb. 2005).

\bibitem{craigie} N.S.~Craigie, Phys. Rept. 47 (1978) 1.

\bibitem{vogt-hard-probes}
% R. Vogt's contribution in 
M. Bedjidian et al.,
% CERN Yellow Book of the Workshop Hard Probes in Heavy Ion Collisions at the LHC,
hep-ph/0311048 (2003).
%\tit{Hard probes in heavy ion collisions at the LHC: heavy flavor physics.}

\bibitem{mrst}
A.D. Martin et al., % R.G. Roberts, W.J. Stirling, R.S. Thorne,
Eur. Phys. J. C28 (2003) 455.
%\tit{``Uncertainties of predictions from parton distributions. 1: experimental errors''.}


\end{thebibliography}
